\documentclass[11pt]{article}
\usepackage{mathrsfs}
\usepackage{graphicx}
\begin{document}
%\begin{CJK*}{GBK}{song}

\title{A New Non-Abelian Topological Phase of Cold Fermi Gases in Anisotropic and Spin-Dependent Optical Lattices }

\author{Beibing Huang \thanks{Corresponding author.
Electronic address: hbb4236@mail.ustc.edu.cn}\\
Department of Experiment Teaching, Yancheng Institute of Technology,
Yancheng, 224051, China
\\Xiaosen Yang and ShaoLong Wan\\
Institute for Theoretical Physics and Department of Modern Physics \\
University of Science and Technology of China, Hefei, 230026, China}

\maketitle
\begin{center}
\begin{minipage}{120mm}
\vskip 0.8in
\begin{center}{\bf Abstract} \end{center}

{To realize non-Abelian s-wave topological superfluid of cold Fermi
gases, generally a Zeeman magnetic field larger than superfluid
pairing gap is necessary. In this paper we find that using an
anisotropic and spin-dependent optical lattice to trap gases, a new
non-Abelian topological superfluid phase appears, in contrast to an
isotropic and spin-independent optical lattice. A characteristic of
this new non-Abelian topological superfluid is that Zeeman magnetic
field can be smaller than the superfluid pairing gap. By
self-consistently solving pairing gap equation and considering the
competition against normal state and phase separation, this new
phase is also stable. Thus an anisotropic and spin-dependent optical
lattice supplies a convenient route to realize topological
superfluid. We also investigate edge states and the effects of a
harmonic trap potential.}

\end{minipage}
\end{center}

\vskip 1cm

\textbf{PACS} number(s): 67.85.Lm, 03.65.Vf, 74.20.-z

\section{Introduction}

In two dimensional condensed matter physics, the search of
non-Abelian Majorana fermions (MF) \cite{majorana}, which are their
own antiparticles and may have potential important application in
fault-tolerant topological quantum computation \cite{sarma},
attracts much recent attention to topological superfluid (TS) and
superconductors (TSC) \cite{qi, hasan}, which has a full pairing gap
in the bulk and topologically protected gapless states on the
boundary. Inspired by the fact that MFs exist at the vortex cores of
a two-dimensional (2D) $p_x+ip_y$ superconductor \cite{read},
theoretically some practical systems have been proposed, such as
fractional quantum Hall effect at filling $\nu=5/2$ \cite{read},
superfluid He-3 \cite{tsut}, non-centrosymmetric superconductors
\cite{sato1, lee}, the surface of a three-dimensional topological
insulator in proximity to an s-wave superconductor \cite{fu, linder}
and a spin-orbit-coupled (SOC) semiconductor quantum well coupled to
an s-wave superconductivity and a ferromagnetic insulator
\cite{jau}. Especially the latter has been improved by applying an
in-plane magnetic field to a (110)-grown semiconductor coupled only
to an s-wave superconductor. This improvement not only enhances the
ability to tune into the topological superconducting state but also
avoid orbital effect of the external magnetic field \cite{anj1}.

It is widely known that cold Fermi gases can be used to simulate
many other systems owing to their many controllable advantages and
operabilities. Certainly the simulations to TS are also possible and
as far as what we know is concerned, three main routes have been
suggested. The first was direct and based on p-wave superfluidity of
degenerate Fermi gases by means of p-wave Feshbach resonance
\cite{gurarie}. Although this method is very simple, it is
challenging due to short lifetimes of the p-wave pairs and
molecules. Subsequently Zhang et al. \cite{zhang} propose to creat
an effective $p_x + ip_y$ TS from an s-wave interaction making use
of an artificially generated SOC. In fact, SOC have been realized in
a neutral atomic Bose-Einstein condensate (BEC) and the same
technique is also feasible for cold Fermi gases \cite{e1, e2}.
Realizing that in a dual transformation SOC is formally equivalent
to a p-wave superfluid gap, Sato et al. \cite{sato} suggest to
artificially generate the vortices of SOC by using lasers carrying
orbital angular momentum. In terms of the latter two ansatzs, in
order to enter into non-Abelian TS, a Zeeman magnetic field larger
than superfluid pairing gap is essentially needed.

In this paper we propose a model to realize non-Abelian TS in cold
Fermi gases by using an anisotropic and spin-dependent optical
lattice (ASDOL) to engineer mismatched Fermi surfaces for each
hyperfine species. Such two requirements for optical lattices are
both accessible in an experiment. The anisotropy is determined by
the intensity of the corresponding pair of laser beams in different
directions, while spin-dependence is also available in the light of
the fact that the strength of the optical potential crucially
depends on the atomic dipole moment between the internal states
involved \cite{mandle1, mandle2, liu, jaksch}. In contrast to an
isotropic and spin-independent optical lattice (ISIOL) \cite{zhang,
sato}, our model realizes another new non-Abelian TS phase in which
Zeeman magnetic field can be smaller than superfluid pairing gap.

The organization of this paper is as follows. In section 2, we
firstly give our model and analyze the condition of gap closing.
Then we calculate TKNN number $I_{TKNN}$ \cite{TKNN} of occupied
bands addressing the topological properties of the model to obtain
topological phase diagram. A new non-Abelian TS is discovered. In
addition we also investigate the properties of edge states to prove
bulk-boundary correspondence. In section 3, by taking the
self-consistency constraint on the s-wave pairing gap into account,
the stability of TS and the effects of a harmonic trap potential are
discussed. We find that this new non-Abelian TS is stable. A brief
conclusion is given in Section 4.

\section{Hamiltonian and Topological Phase Diagram}

In this paper we consider s-wave superfluid of cold Fermi gases with
Rashba SOC in a two-dimension square optical lattice which is
anisotropic and spin-dependent. For convenience below we assume the
lattice constant to be unit. The Hamiltonian suggested is
\begin{eqnarray}
H=\sum_{k\sigma}[\epsilon_{k\sigma}-\mu-\sigma\Gamma]a_{k\sigma}^{\dag}a_{k\sigma}+\sum_k[
J_k a_{k\uparrow}^{\dag}a_{k\downarrow}-\Delta
a_{-k\downarrow}a_{k\uparrow}+H.C.],\label{1}
\end{eqnarray}
where $a_{k\sigma}^{\dag}$ creates a spin
$\sigma=\uparrow,\downarrow$ fermion at momentum $\vec{k}=(k_x,
k_y)$. $J_k=2J[\sin{(k_y)}+i \sin{(k_x)}]$ with $J$ ($J>0$) denoting
the strength of Rashba SOC. $\mu$, $\Gamma$, $\Delta$ are chemical
potential, effective Zeeman magnetic field and s-wave superfluid
pairing gap, respectively.

The kinetic energy terms $\epsilon_{k\sigma}$ come from anisotropic
and spin-dependent optical lattices \cite{fisher}. Imagine tuning
the intensities of lasers so that one spin state prefers to hop
along the $x$ axis and the other prefers to hop along the $y$ axis.
In this paper we concentrate on a specific situation where the Fermi
surfaces of the two spin states are rotated by $90^0$ with respect
to one another, and only consider a near-neighbor hopping
Hamiltonian with single particle dispersions
\begin{eqnarray}
\epsilon_{k\uparrow}=-2t_a\cos{(k_x)}-2t_b\cos{(k_y)}, \nonumber\\
\epsilon_{k\downarrow}=-2t_b\cos{(k_x)}-2t_a\cos{(k_y)}. \label{2}
\end{eqnarray}

When there are no Zeeman magnetic field and SOC $\Gamma=J=0$, this
model realized a stable paired superfluid state with coexisting
pockets of momentum space with gapless unpaired fermions
\cite{fisher}, similar to the Sarma state in polarized mixtures
\cite{sar, liuw, forbes}. Moreover in the strong coupling limit, a
d-wave pairing superfluid as well as a d-wave density wave state,
are also proposed to be achievable in this system \cite{caizi}.

Generally speaking, the close of bulk gap is a signal of topological
phase transitions, although it is not a sufficient condition
\cite{law}. Thus to obtain topological phase diagram of Hamiltonian
(\ref{1}), we firstly calculate the bulk spectrum of the system to
search parameter regions for which different topological phases are
possible. Then for every regions we calculate topological invariants
to label topological properties. To obtain the bulk spectrum, we
rewrite the Hamiltonian as
\begin{eqnarray}
H=\frac{1}{2}\sum_k\psi_{k}^{\dag}M_{4\times4}\psi_{k},\label{3}
\end{eqnarray}
where $\psi_{k}^{\dag}=(a_{k\uparrow}^{\dag},a_{k\downarrow}^{\dag},
a_{-k\uparrow},a_{-k\downarrow})$ is a row vector and $M_{4\times4}$
is a matrix with $M_{11}=-M_{33}=\epsilon_{k\uparrow}-\mu-\Gamma$,
$M_{22}=-M_{44}=\epsilon_{k\downarrow}-\mu+\Gamma$,
$M_{12}=M_{43}=M_{21}^{\ast}=M_{34}^{\ast}=J_k$,
$M_{23}=M_{32}=-M_{14}=-M_{41}=\Delta$,
$M_{13}=M_{31}=M_{24}=M_{42}=0$.

Diagonalizing the matrix $M_{4\times4}$, we find the energy spectrum
\begin{eqnarray}
E_k^{\pm}=\sqrt{\xi_{k+}^2+\xi_{k-}^2+|J_k|^2+\Delta^2 \pm 2E_0}
\label{4}
\end{eqnarray}
with $\xi_{k+}=-(t_a+t_b)[\cos{(k_x)}+\cos{(k_y)}]-\mu$,
$\xi_{k-}=(-t_a+t_b)[\cos{(k_x)}-\cos{(k_y)}]-\Gamma$ and $E_0=
\sqrt{\xi_{k+}^2\xi_{k-}^2+\xi_{k+}^2|J_k|^2+\xi_{k-}^2\Delta^2}$.

The close of the bulk energy gap is possible only if $E_k^-=0$, in
other words
\begin{eqnarray}
\xi_{k+}^2+\xi_{k-}^2+|J_k|^2+\Delta^2=2E_0, \label{5}
\end{eqnarray}
which is equivalent to
\begin{eqnarray}
\xi_{k+}^2-\xi_{k-}^2-|J_k|^2+\Delta^2=0,\quad\quad
|J_k|^2\Delta^2=0.\label{6}
\end{eqnarray}

For the s-wave pairing,  $\Delta\neq 0$ and the second equation in
(\ref{6}) is satisfied only when $k = (0, 0)$, $(0, \pi)$, $(\pi,
0)$, $(\pi, \pi)$. Substituting these values into the first equation
in (\ref{6}), four different gap closing conditions are obtained
\begin{eqnarray}
\mu^2+\Delta^2=[2(t_b-t_a)+\Gamma]^2,\quad
\mu^2+\Delta^2=[2(t_b-t_a)-\Gamma]^2,\quad
\nonumber\\\Gamma^2=\Delta^2+ [2(t_a+t_b)+\mu]^2,\quad
\Gamma^2=\Delta^2+ [2(t_a+t_b)-\mu]^2.\label{7}
\end{eqnarray}
By means of these conditions, Fig.1(a) shows that there are at least
$12$ regions, which may be topologically distinct. By now we have
completed the first step of deciding topological phase diagram.
Below we will explore the topological numbers to classify the
topological phases of the model (\ref{1}).

In terms of our system, it explicitly breaks the time-reversal
symmetry even though Zeeman magnetic field is zero. Thus TKNN number
$I_{TKNN}$ plays a central role in topological nature of the system
\cite{TKNN}, which is consistent with the conclusion in the periodic
table of TSC that BdG Hamiltonian only with particle-hole symmetry
in two dimension is classified by integer group $Z$ \cite{ludwig}.
Furthermore the topological properties of the system are identified
as follows \cite{satom}. If TKNN number is nonzero and even (odd),
the system is Abelian (non-Abelian) TS without (with) non-Abelian
anyons; If TKNN number is zero, the system is topologically trivial
and is normal superfluid.

TKNN number is defined by
\begin{eqnarray}
I_{TKNN}=\frac{1}{2\pi i}\int d^2k F_{12}(k),\label{8}
\end{eqnarray}
where Berry connection $A_{i}(k)$ ($i=1,2$) and associated field
strength $F_{12}(k)$ are given by
\begin{eqnarray}
A_i(k)&=&\sum_{E_n<0}<\phi_n(k) |\partial_{k_i}\phi_n(k)>\nonumber\\
F_{12}(k)&=&\partial_1 A_2(k) - \partial_2 A_1(k) \label{9}
\end{eqnarray}
with $|\phi_n(k)>$ being a normalized wave function of the $n$th
band such that $M_{4\times4}|\phi_n(k)>=E_n(k)|\phi_n(k)>$. It is
should be noted that the sum in (\ref{9}) is restricted to occupied
bands.

TKNN numbers of $12$ regions have been numerically calculated
\cite{fukui} and given in Fig.1(a). Totally $5$ non-Abelian TS and
one Abelian TS phases are found. In order to make a contrast, in
Fig.1(b) we also show the topological phase diagram of an ISIOL and
there are 4 non-Abelian TS and 1 Abelian TS phases. By comparison
the effects of ASDOL are mainly twofolds. On the one hand it creates
another non-Abelian TS phase around the chemical potential $\mu= 0$,
which can be realized for Zeeman magnetic field $\Gamma$ less than
superfluid pairing gap $\Delta$. This is our main result in this
paper and can be understood from the fact that an ASDOL ($t_a\neq
t_b$) supplies an effective Zeeman magnetic field $\pm2(t_a-t_b)$,
as seen from (7). On the other hand in the direction of increasing
Zeeman magnetic field it separates two successive non-Abelian TS
phases in Fig.1(b).

According to the bulk-boundary correspondence, a topologically
nontrivial bulk guarantees the existence of topologically stable
gapless edge states on the boundary. Cold Fermi gases with sharp
edges may be realized along the lines proposed in \cite{goldman}. In
our case, the gapless edge states is a chiral Majorana fermion mode.
It should also be remembered that the core of a vortex is
topologically equivalent to an edge which has been closed on itself.
The topologically protected edge modes we describe here are
therefore equivalent to the Majorana fermions known to exist in the
core of vortices of p-wave superfluid \cite{lewenstein}.

To study edge states, we transform Hamiltonian (\ref{1}) into
lattice representation
\begin{eqnarray}
H&=&-t_a\sum_i [ a_{i\uparrow}^{\dag}a_{i+x\uparrow}+
a_{i\downarrow}^{\dag}a_{i+y\downarrow}+H.C.] -t_b\sum_i [
a_{i\uparrow}^{\dag}a_{i+y\uparrow}+
a_{i\downarrow}^{\dag}a_{i+x\downarrow}+H.C.]\nonumber\\
&&+J\sum_i[(a_{i\uparrow}^{\dag}a_{i+x\downarrow}-a_{i\downarrow}^{\dag}a_{i+x\uparrow})
-i(a_{i\uparrow}^{\dag}a_{i+y\downarrow}+a_{i\downarrow}^{\dag}a_{i+y\uparrow})+H.C.]\nonumber\\
&&-\sum_{i\sigma}(\mu+\sigma\Gamma)a_{i\sigma}^{\dag}a_{i\sigma}-
\Delta\sum_i(a_{i\uparrow}^{\dag}a_{i\downarrow}^{\dag}+H.C.)
\label{10}
\end{eqnarray}
where $a_{i\sigma}^{\dag}$ is the creation operator of a fermion
with spin $\sigma$ at lattice site $i= (i_x, i_y)$. Without loss of
generality we suppose that the system has two open boundary in the
$x$-direction and is periodic in the $y$-direction. After performing
a Fourier transformation along the $y$-direction only, we
numerically diagonalize the Hamiltonian (\ref{10}) and
correspondingly obtain the excitation spectrum $E_{n,k_y}$ with
subscript $n$ labeling different energy levels.

In Fig.2 we shows the energy spectra for an ASIOL with edges at $i_x
= 0$ and $i_x = 30$. These $12$ figures correspond to $12$ regions
in Fig.1(a), respectively. Fig.2(b), (d), (f), (h), (j) are for
non-Abelian TS phases, (k) for Abelian TS phase and others non-TS
phases. As stated in \cite{satom, lewenstein}, we also find that in
the non-Abelian TS phases, where $I_{TKNN}=\pm 1$ and
$(-1)^{I_{TKNN}}=-1$, a single pair of gapless edge modes appears.
Since on a given boundary there is no state available for backward
spinconserving scattering, these states are topologically protected
\cite{lewenstein}. In this case the edge zero mode is a chiral
Majorana fermion. While in Abelian TS phase $I_{TKNN}=2$ and non-TS
phases $I_{TKNN}=0$, where $(-1)^{I_{TKNN}}=1$, the system contains
either zero or two pairs of edge states. These edge states are
topologically trivial for either edge modes can perform
spin-conserving backscattering. In addition it is also found that
edge excitations only cross at zero energy with a linear dispersion
at $k_y = 0$ or $\pi$. Their origin is topological and can be
identified by a topological winding number $I(k_y)$ defined only for
$k_y = 0$ and $\pi$ \cite{satom}: when $I(k_y)$ is non-zero for $k_y
= 0$ or $\pi$, the energy of the gapless edge mode becomes zero at
this value of $k_y$.

\section{Phase Diagram with Self-Consistent Pairing Gap}

In section 2 we have determined the topological phase diagram for a
fixed s-wave superfluid pairing gap. However for a realistic
physical system the pairing gap generally varies when the chemical
potential and Zeeman magnetic field change. Especially Zeeman
magnetic field breaks time-reversal symmetry and weakens the
stability of superfluidity. A well-known example is so-called
Chandrasekar-Clogston (CC) limit \cite{cc, cc1} in superconducting
systems without SOC. Hence it is possible that not all phases in
Fig.1(a) are accessible. In this section within BCS mean-field
theory, we self-consistently determine s-wave superfluid pairing gap
and consider the competition from normal phase and phase separation
to investigate the stability of TS.

Let $-U$ ($U>0$) denote the effective attraction strength between
fermions, then the pairing gap
$\Delta=U\sum_k<a_{-k\downarrow}a_{k\uparrow}>$ can be obtained from
the minimization of thermodynamic potential $\Omega_s=\sum_k
\left[\xi_{k+}-\frac{1}{2}(E_k^{+}+E_k^{-})\right]+N \Delta^2/U$.
The unstability of superfluidity against phase separation is
signalled by the condition $\Delta\neq 0$ and $\partial^2
\Omega_s/\partial \Delta^2<0$, while unstability against normal
state is $\Omega_n<\Omega_s$, where $\Omega_n$ is thermodynamic
potential of normal state. For parameter chosen superfluidity is
robust as long as $\Delta\ne0$. The resulting zero-temperature
pairing gap and phase diagram are shown in Fig.3. Fig.3(b) shows
similar structures with Fig.1(a) except TS with large Zeeman
magnetic field are not available. Comparing Fig.3(a) with (b), it is
easily seen that for small chemical potential and Zeeman magnetic
field, although the pairing gap has a larger magnitude ($\Delta\sim
t_a$) than Zeeman magnetic field, the system can be non-Abelian TS
with $I_{TKNN}=-1$. This finding is very important and suggests that
such a new non-Abelian TS we have found in section 2 is stable.

At last we consider the effects of a harmonic trap potential by
local density approximation (LDA). Under this approximation the
system is locally uniform and trap potential effectively provides a
scan over the chemical potential, in other words local chemical
potential $\mu_R=\mu-\Delta_t R^2$ with $\Delta_t$ denoting the
strength of an isotropic potential. From Fig.3(b) depending on
Zeeman magnetic field some different topological phases can coexist
in different concentric spherical shells. In Fig.4 we show the space
distributions of pairing gap, particle number and TKNN number. It is
should be noted that on the boundary of different topological phases
there always exists zero-energy states due to the change of
topological invariant.

\section{Conclusions}

In conclusion we have investigated the effects of an anisotropic and
spin-dependent optical lattice (ASDOL) on non-Abelian topological
superfluid and found that a new non-Abelian topological superfluid
phase exists steadily, in contrast to an isotropic and
spin-independent optical lattice. Moreover in this new non-Abelian
topological superfluid phase Zeeman magnetic field can be smaller
than the superfluid pairing gap. Thus an ASDOL supplies a convenient
route to realize TS. In addition we also calculated chiral Majorana
edge states and investigated the effects of a harmonic trap
potential.

\section*{Acknowledgement}

The work was supported by National Natural Science Foundation of
China under Grant No. 10675108 and Foundation of Yancheng Institute
of Technology under Grant No. XKR2010007.

\begin{figure}[htbp]
\includegraphics[width=7.5cm, height=6.0cm]{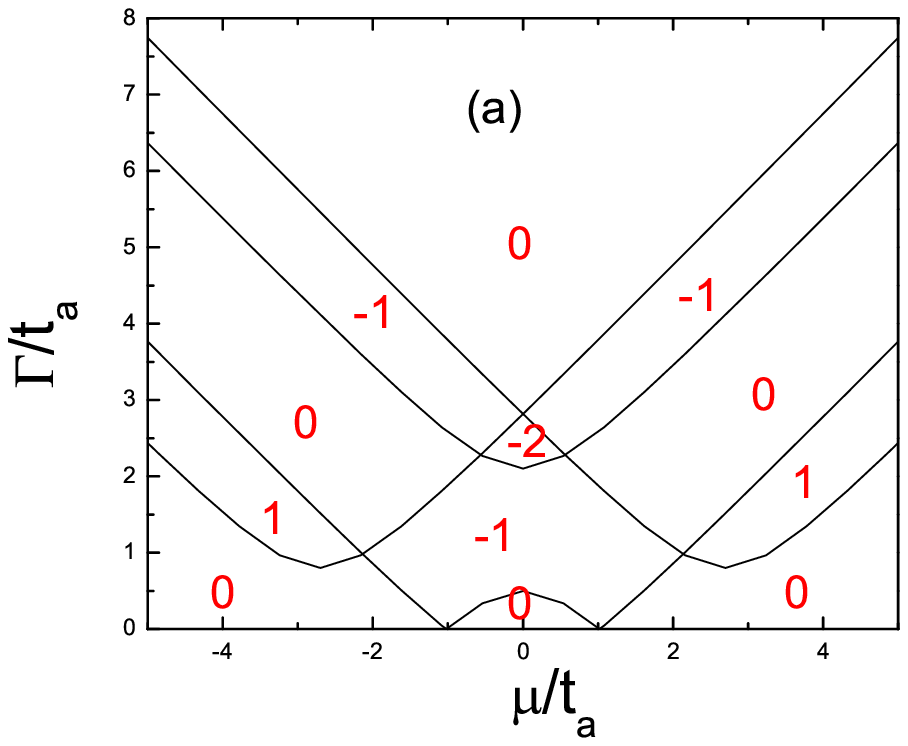}
\includegraphics[width=7.5cm, height=6.0cm]{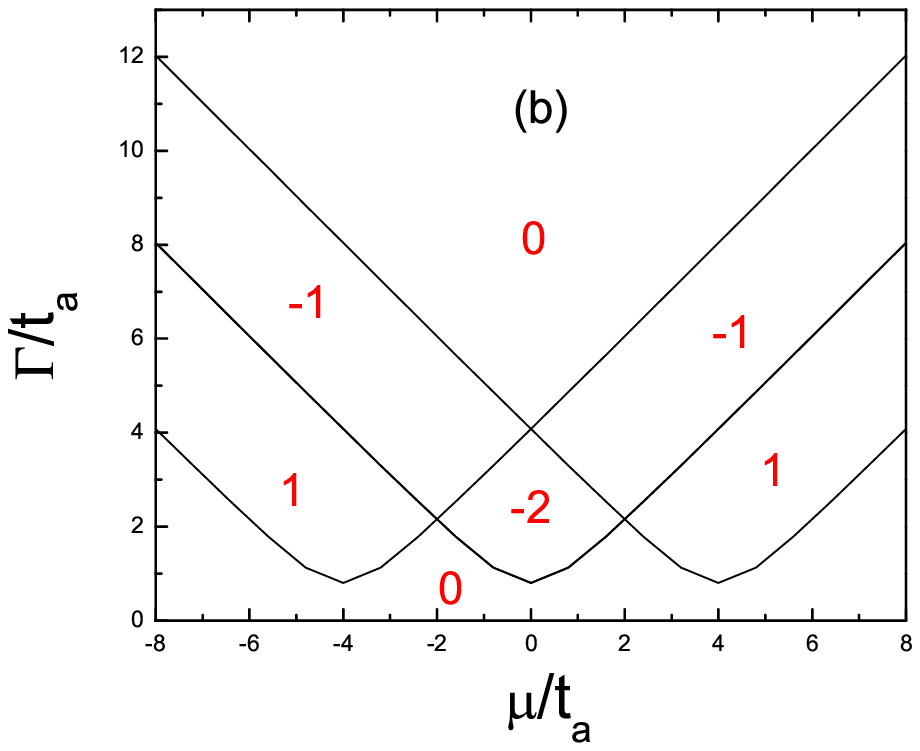}
\caption{Topological phase diagram. The numbers in different regions
are TKNN numbers. We have chosen $t_b/t_a=0.35$, $J/t_a=0.5$ and
$\Delta/t_a=0.8$. (a) is for an anisotropic and spin-dependent
optical lattice and (b) is for an isotropic and spin-independent
optical lattice.} \label{fig.1}
\end{figure}

\begin{figure}[htbp]
\includegraphics[width=5.0cm, height=4.0cm]{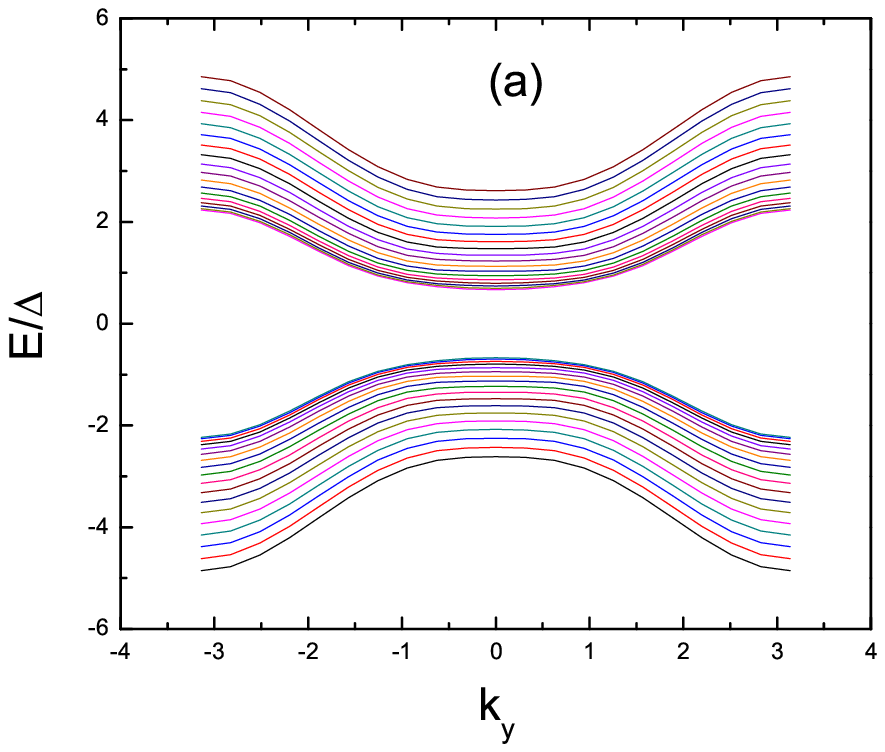}
\includegraphics[width=5.0cm, height=4.0cm]{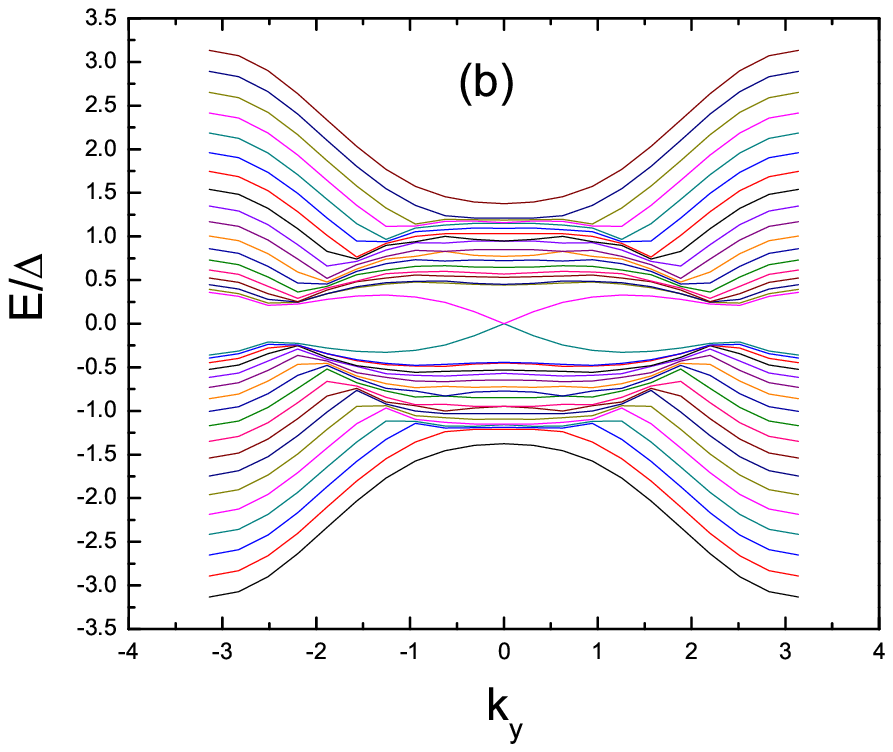}
\includegraphics[width=5.0cm, height=4.0cm]{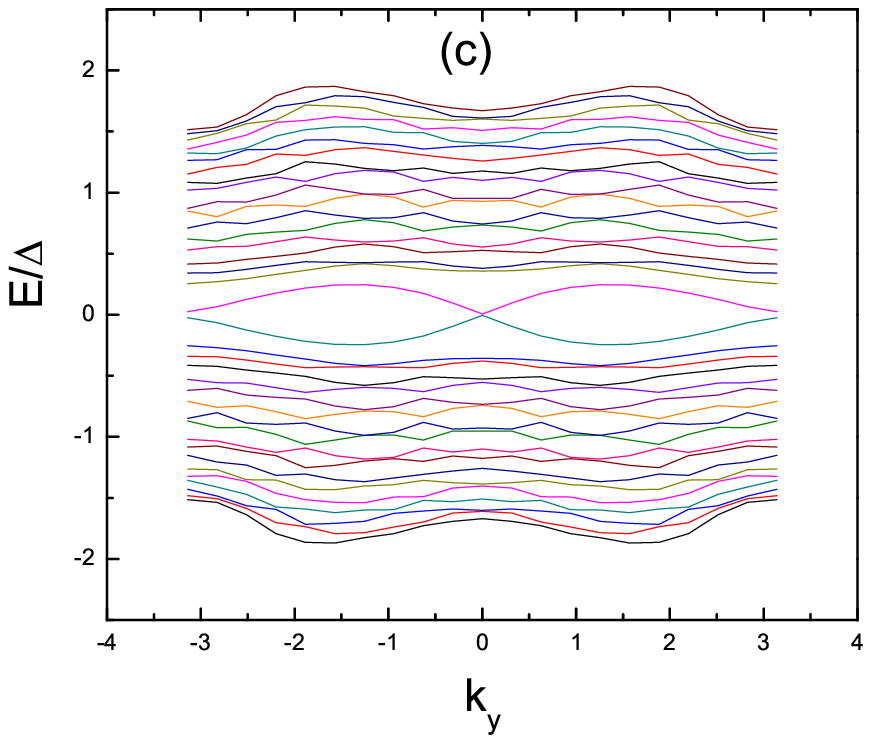}
\includegraphics[width=5.0cm, height=4.0cm]{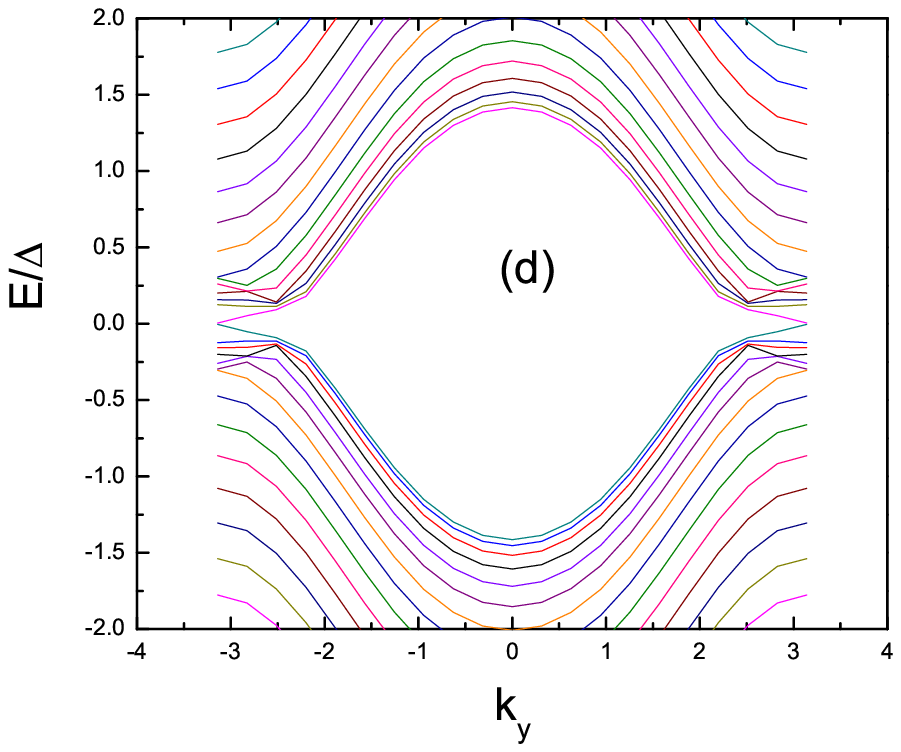}
\includegraphics[width=5.0cm, height=4.0cm]{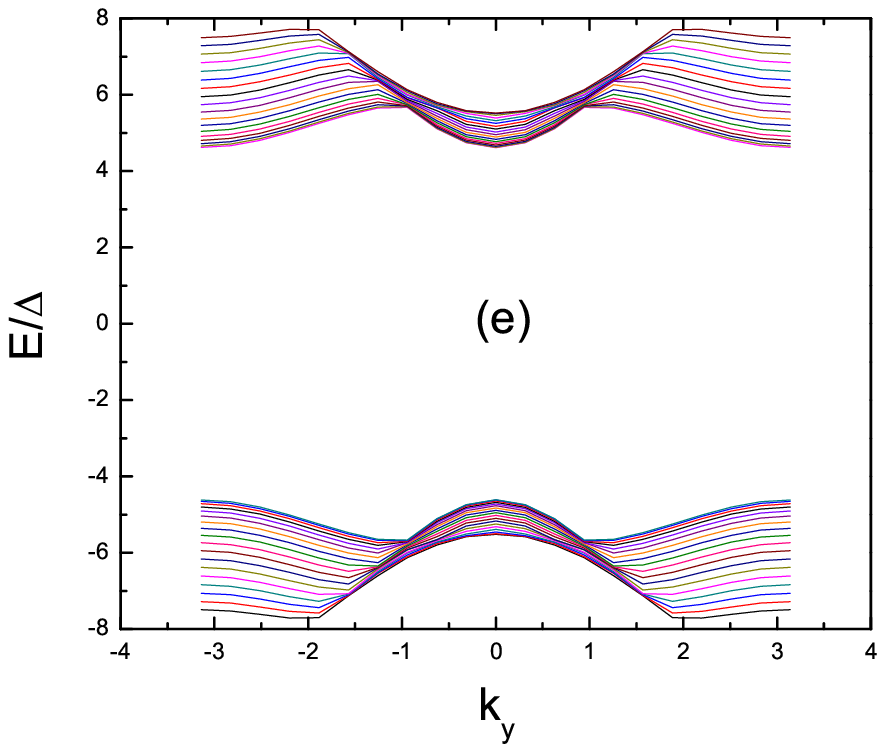}
\includegraphics[width=5.0cm, height=4.0cm]{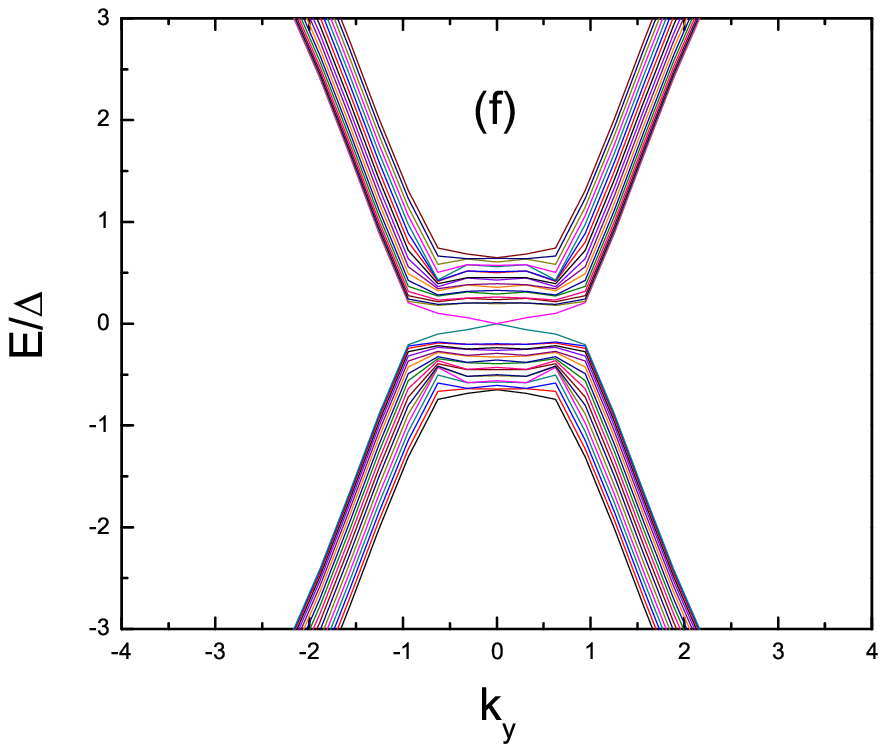}
\includegraphics[width=5.0cm, height=4.0cm]{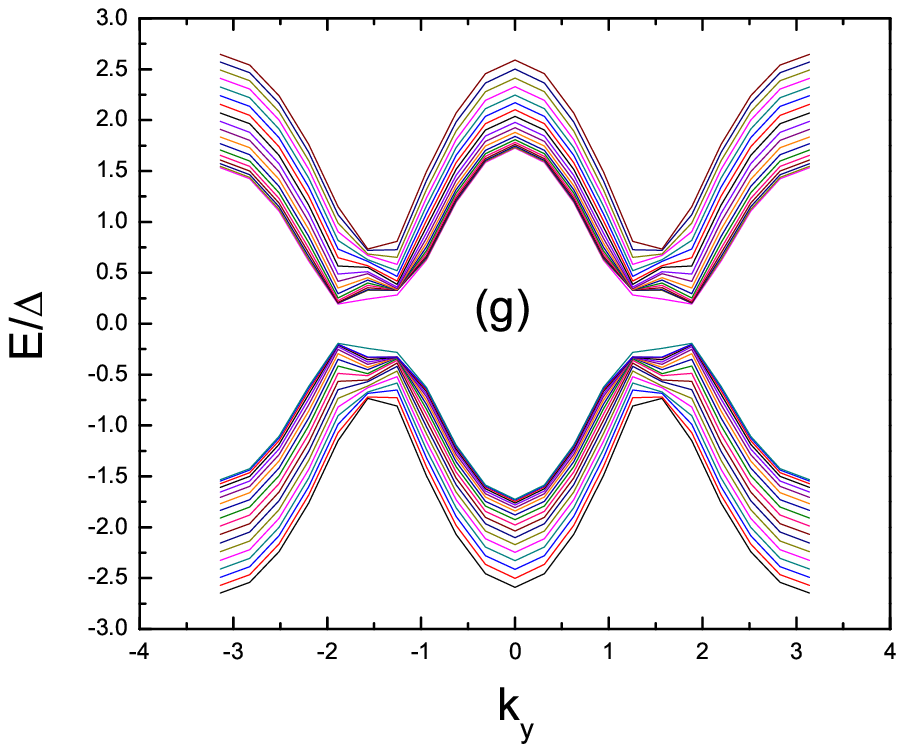}
\includegraphics[width=5.0cm, height=4.0cm]{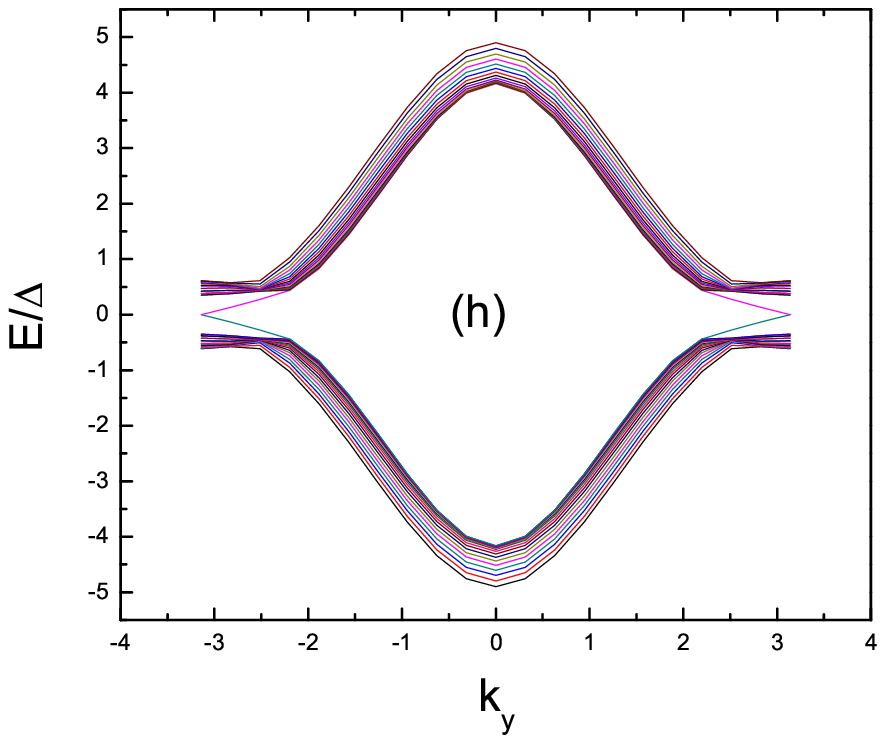}
\includegraphics[width=5.0cm, height=4.0cm]{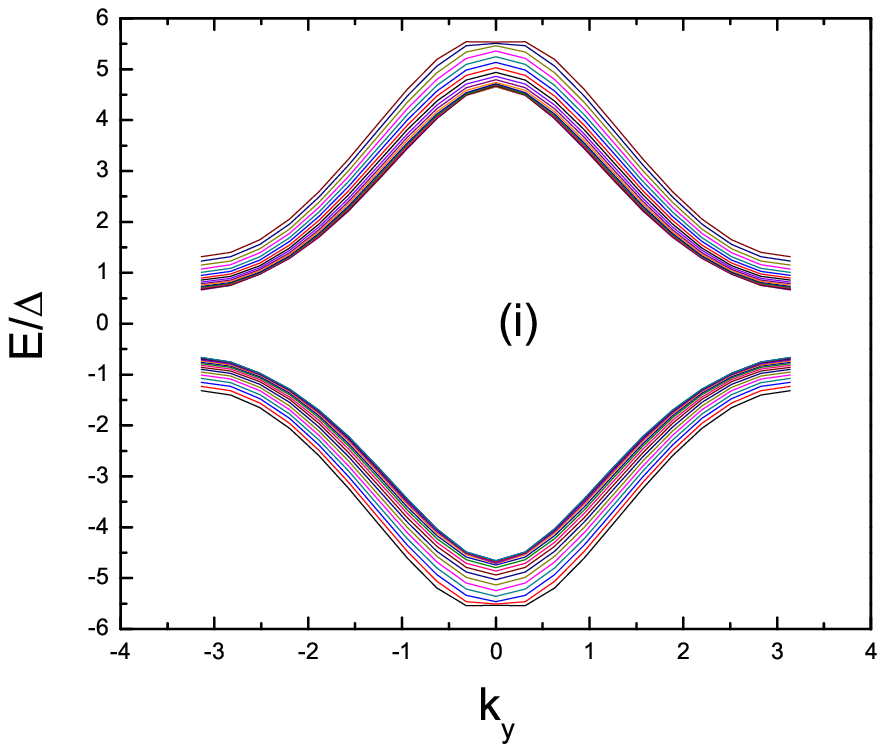}
\includegraphics[width=5.0cm, height=4.0cm]{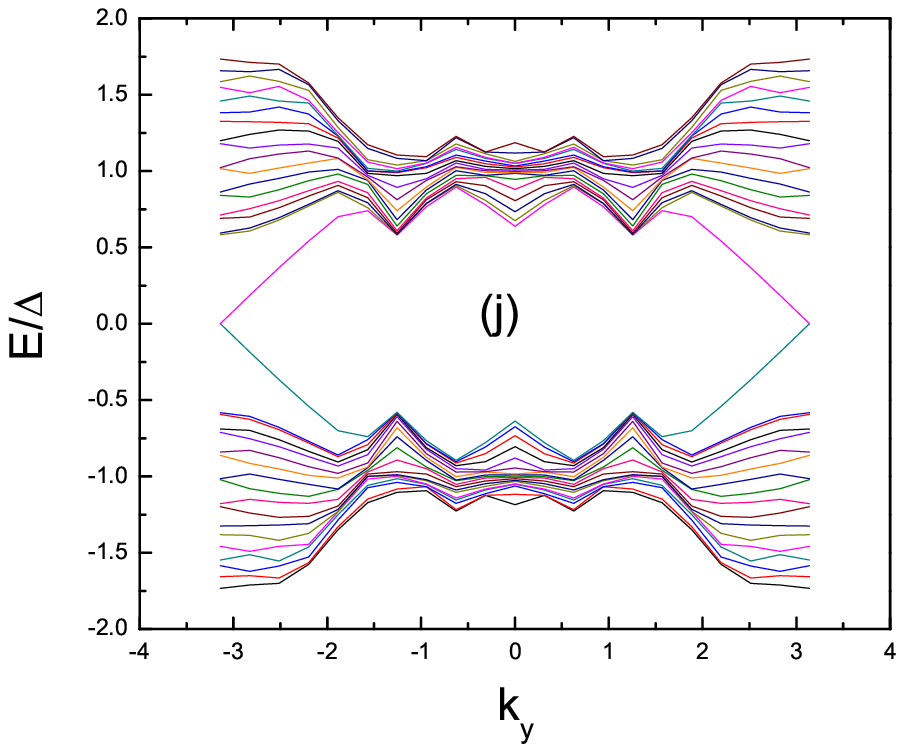}
\includegraphics[width=5.0cm, height=4.0cm]{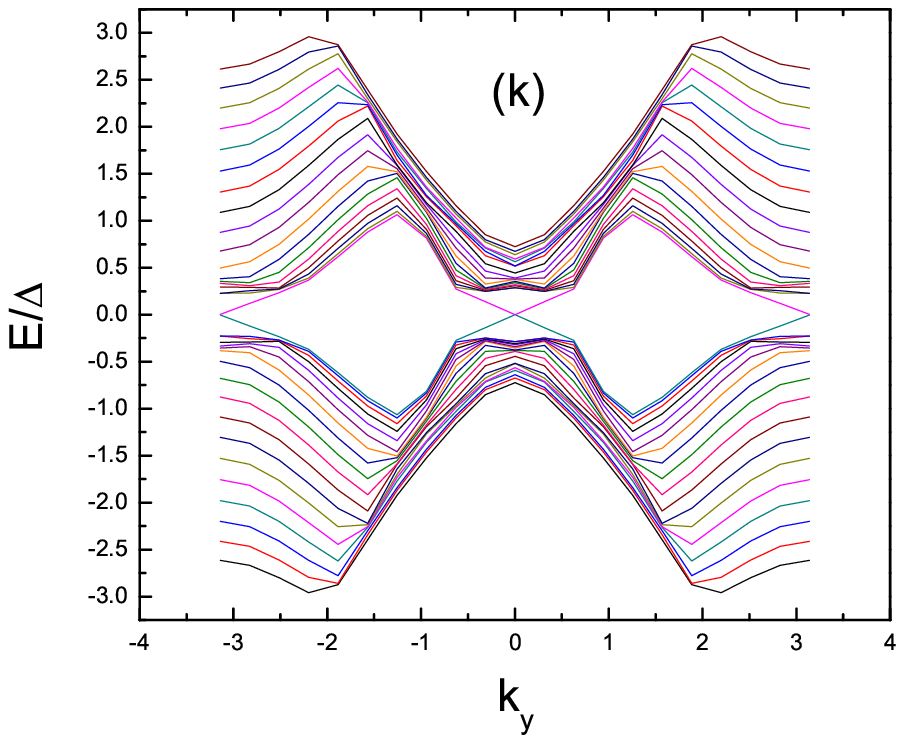}
\includegraphics[width=5.0cm, height=4.0cm]{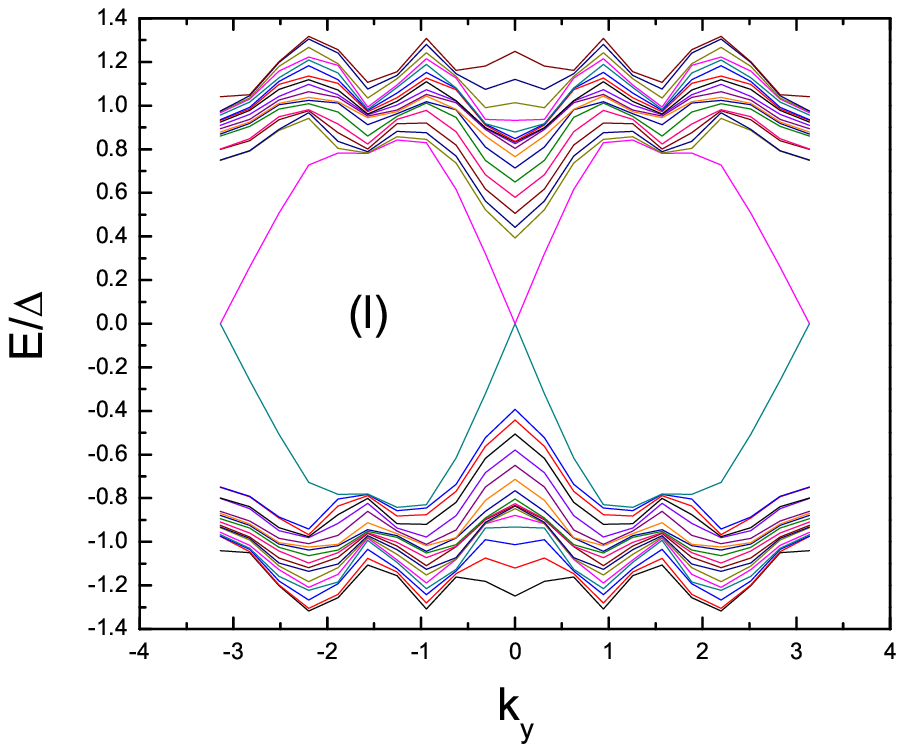}
\caption{The energy spectra of an anisotropic and spin-dependent
optical lattice with open edges at $i_x = 0$ and $i_x = 30$. We have
chosen $t_b/t_a=0.35$, $J/t_a=0.5$ and $\Delta/t_a=0.8$. (a)
$\mu/t_a=-4.0$, $\Gamma/t_a=1.0$, (b) $\mu/t_a=-4.0$,
$\Gamma/t_a=2.5$, (c) $\mu/t_a=-4.0$, $\Gamma/t_a=4.0$, (d)
$\mu/t_a=-4.0$, $\Gamma/t_a=6.5$, (e) $\mu/t_a=0.0$,
$\Gamma/t_a=6.5$, (f) $\mu/t_a=4.0$, $\Gamma/t_a=6.0$, (g)
$\mu/t_a=4.0$, $\Gamma/t_a=4.0$, (h) $\mu/t_a=4.0$,
$\Gamma/t_a=2.0$, (i) $\mu/t_a=4.0$, $\Gamma/t_a=1.0$, (j)
$\mu/t_a=0.0$, $\Gamma/t_a=1.0$, (k) $\mu/t_a=0.0$,
$\Gamma/t_a=2.5$, (l) $\mu/t_a=0.0$, $\Gamma/t_a=0.2$. These figures
correspond to different regions in Fig.1(a), respectively.}
\label{fig.2}
\end{figure}

\begin{figure}[htbp]
\includegraphics[width=7.5cm, height=6.0cm]{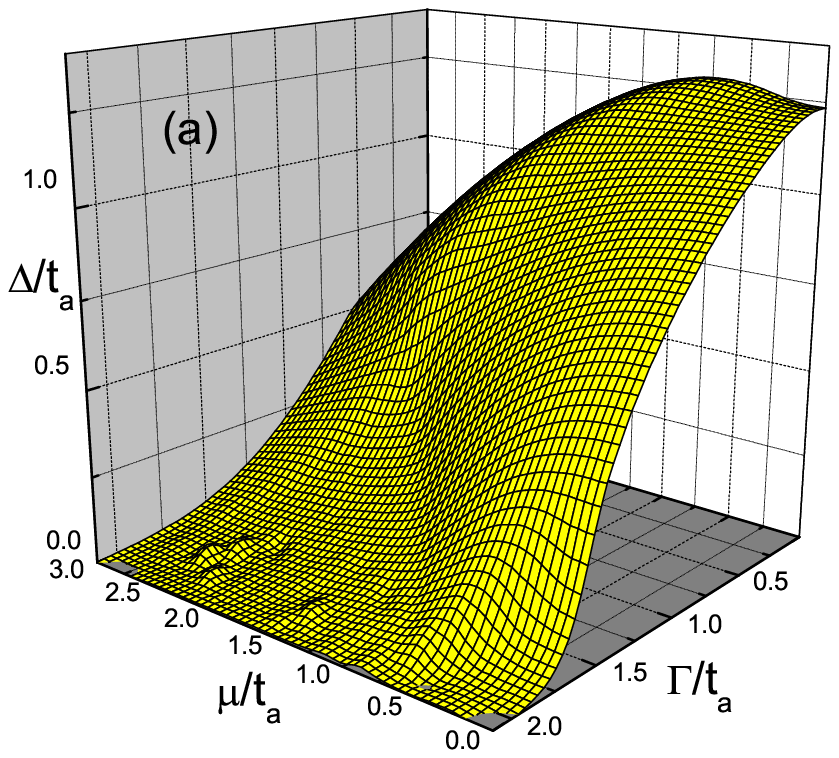}
\includegraphics[width=7.5cm, height=6.0cm]{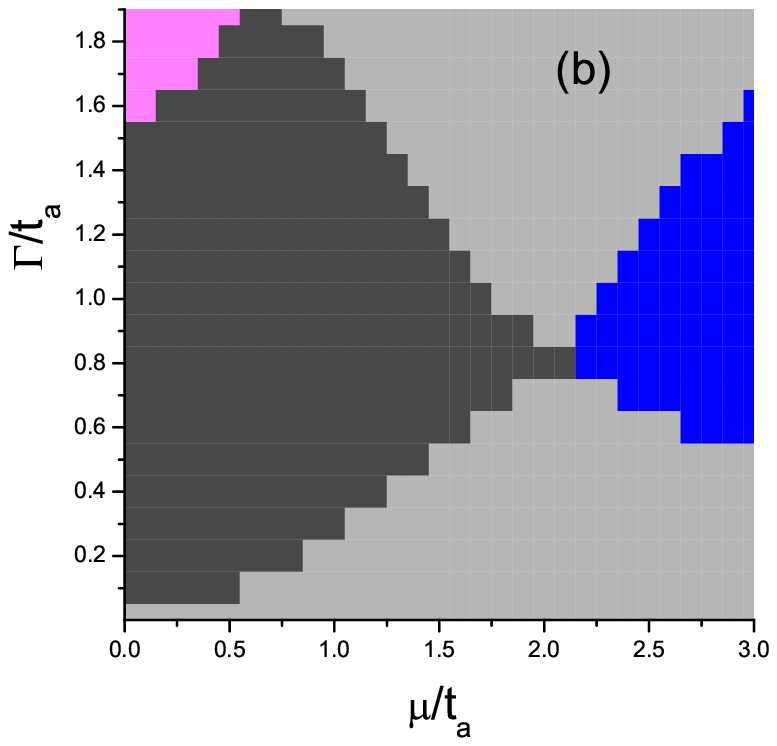}
\caption{s-wave superfluid pairing gap (a) and topological phase
diagram (b) from self-consistent mean-field solution. In (b) grey,
black, blue and purple colors correspond to $I_{TKNN}=0,-1,1,-2$
respectively. We have chosen $t_b/t_a=0.35$, $J/t_a=0.5$ and
$U/t_a=4.0$. } \label{fig.3}
\end{figure}

\begin{figure}[htbp]
\includegraphics[width=7.5cm, height=6.0cm]{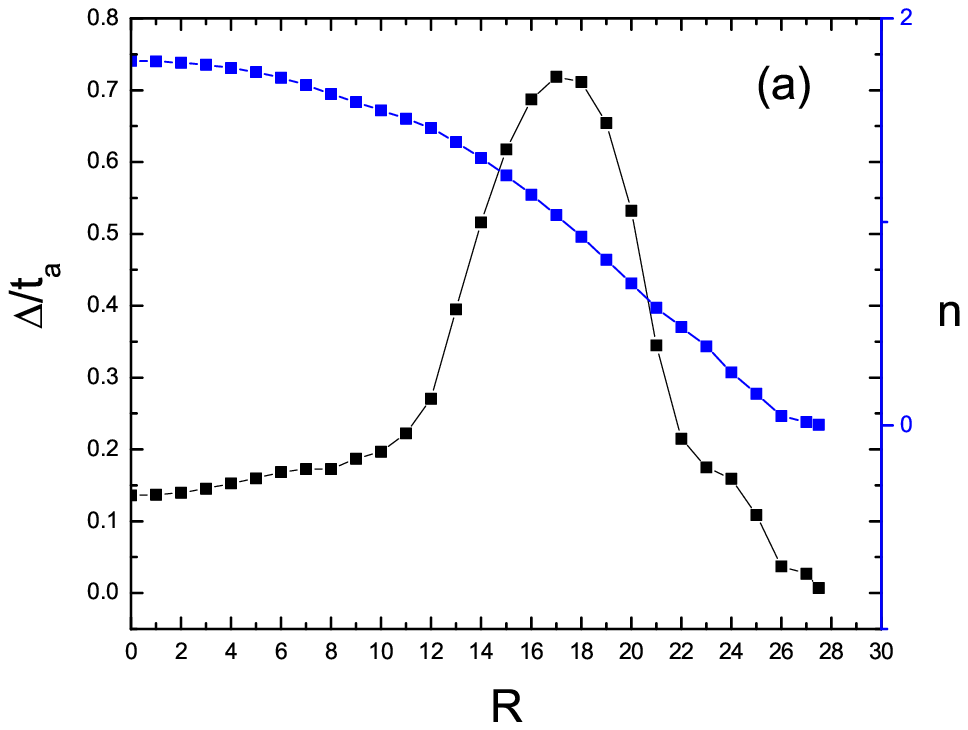}
\includegraphics[width=7.5cm, height=6.0cm]{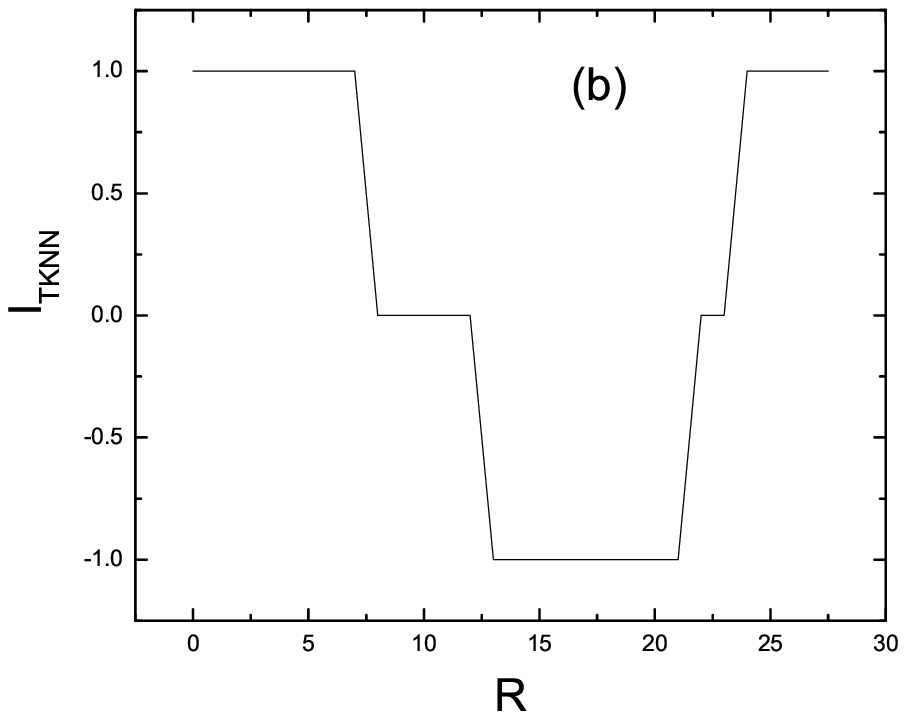}
\caption{The space distributions of pairing gap, particle number and
TKNN number. We have chosen $t_b/t_a=0.35$, $J/t_a=0.5$,
$\mu/t_a=3$, $\Delta_t/t_a=0.01$, $\Gamma/t_a=1.2$ and $U/t_a=4.0$.
} \label{fig.4}
\end{figure}

%\end{CJK*}

\end{document}